\documentclass[review]{elsarticle}
\usepackage{color}
\usepackage{amssymb}
\usepackage{amsmath}
\usepackage{lineno,hyperref}
\usepackage{float}
\modulolinenumbers[5]

\journal{Journal of \LaTeX\ Templates}

\begin{document}

\begin{frontmatter}
%\addbibresource{literature.bib}

\title{Role of rapidity choice for the impact-parameter dependent Balitsky-Kovchegov equation}

\author[CVUT]{Matej Vaculciak (\textit{matej.vaculciak@fjfi.cvut.cz})}
\author[CVUT]{Jesus Guillermo Contreras}
\author[CVUT]{Jan Cepila}
\address[CVUT]{Faculty of Nuclear Sciences and Physical Engineering,
Czech Technical University in Prague, Czech Republic}

\begin{abstract}
Reaching higher energies of electron-ion collisions with facilities like EIC is expected to provide a probe of a kinematic region where the parton densities should start to exhibit signs of saturation.
This phenomenon is theoretically implemented by the Balitsky--Kovchegov (BK) equation, which, within the colour dipole model, describes the evolution of the dipole scattering amplitude with respect to rapidity. There are two possible formulations of the BK equation based on which rapidity, projectile or target, is considered.
Besides this variable, there are four more degrees of freedom, two of which have been so far incorporated into the numerical solutions.
We present a comparative solution of the two-dimensional BK equation formulated in both projectile and target rapidity together with their impact on quantities to be observed at EIC such as proton structure functions.
\newline
The contribution was presented at the Hot Quarks 2022 - Workshop for young scientists on the physics of ultrarelativistic nucleus-nucleus collisions, Dao House, Colorado, USA, October 11-17 2022.
\end{abstract}

\begin{keyword}
QCD phenomenology; parton saturation; Balitsky--Kovchegov equation; low-$x$ QCD
\end{keyword}

\end{frontmatter}

%%%%%%%%%%%%%%%%%%%%%%%%%%%%%%%%%%%%%%%%%%

\section{Introduction}
Theoretical studies of the hadron structure resulted in the need to restrict the growth of parton densities at low Bjorken-$x$, so that the Froissart bound is not violated. Such saturation mechanism can be implemented by the non-linear Balitsky-Kovchegov (BK) evolution equation. See for example Ref.~\cite{Kovchegov:2012mbw}.

Experimentally, deeply inelastic electron-proton scattering is used to study the hadron structure. The leading order of this interaction is mediated by the exchange of a virtual photon, whose emission from the electron is a simple process to describe within quantum electrodynamics. On the other hand, the description of the effective vertex between the proton and the virtual photon is non-trivial and is treated within the colour dipole model \cite{Kovchegov:2012mbw}. Here the virtual photon is replaced with its simplest strongly interacting Fock state---the colour dipole---and the interaction between the dipole and the proton is then described by the so-called dipole scattering amplitude $N(Y, \vec{r}, \vec{b})$. This object depends on the rapidity $Y$ and two two-dimensional vectors describing the dipole size $(\vec{r})$ and its transverse position with respect to the proton $(\vec{b})$. 

To calculate the dipole scattering amplitude, the BK equation 
\begin{align}
\partial_Y N(\vec{r}, \vec{b}, Y) = \int \mathrm{d} \vec{r_1} K(r, r_1, r_2) \Big[ 	&   N(\vec{r}_1, \vec{b}_1, Y) + N(\vec{r}_2, \vec{b}_2, Y) - N(\vec{r}, \vec{b}, Y)	\nonumber \\
		& - N(\vec{r}_1, \vec{b}_1, Y) N(\vec{r}_2, \vec{b}_2, Y) \Big], \label{eq_bk_p}
\end{align}
is used. It is an integro--differential equation and in the current models, it is formulated in terms of the rapidity of the projectile (the virtual photon), defined as
\begin{align}
    Y = \ln{\frac{x_0}{x}},
\end{align}
with $x$ being the Bjorken-$x$ and $x_0$ a parameter defining the evolution starting point.

%%%%%%%%%%%%%%%%%%%%%%%%%%%%%%%%%%%%%%%%%%

\section{BK equation in projectile rapidity}
Due to the complexity of the BK equation (\ref{eq_bk_p}), only numerical solutions have been obtained. To date a solution incorporating the dependence on all four integration variables $\vec{r}, \Vec{b}$ (the 4D solution) is not yet known. 
%However, a 
So far, the state of the art is to find 2D solutions depending on the norms of the dipole size and the impact parameter $N(Y, r, b)$; results have been presented in Ref.~\cite{Cepila:2019fiz} exhibiting a reasonable agreement with experimental data.

\begin{figure}[!ht]
\centering
\includegraphics[width=.49\linewidth]{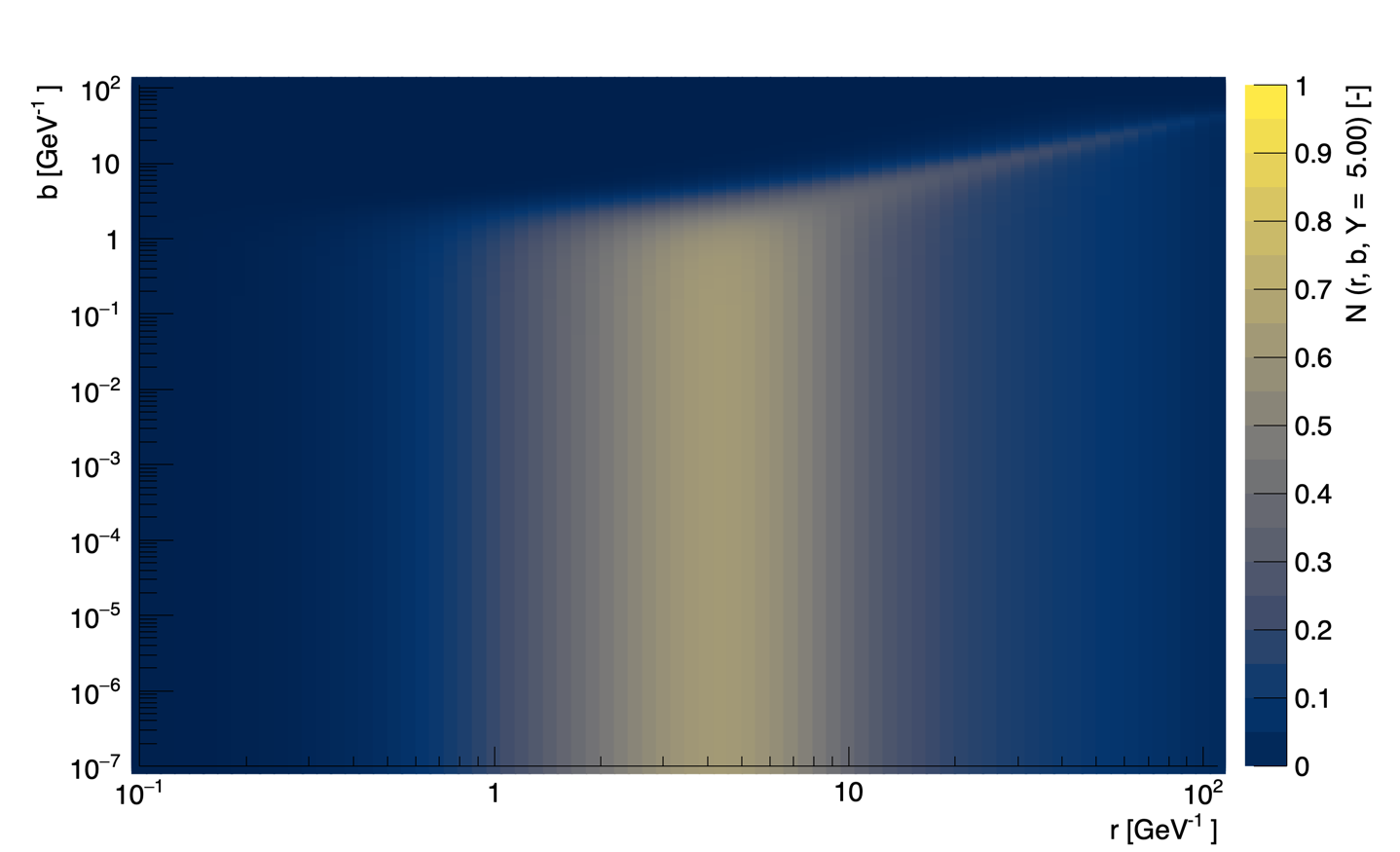}
\includegraphics[width=.49\linewidth]{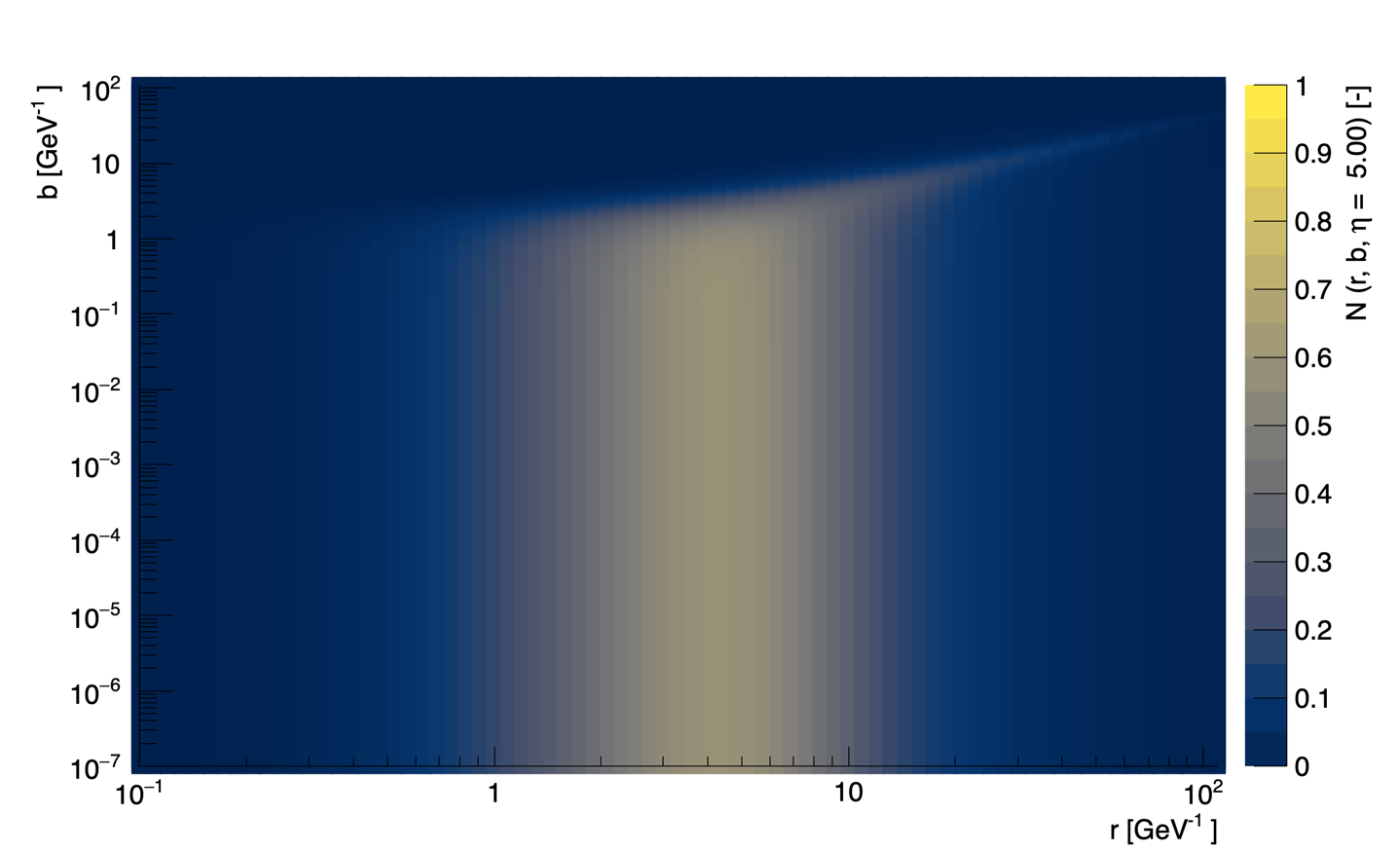}
\caption{
The dipole scattering amplitude evolved to rapidity $Y = \eta = 5$ using the BK equation in projectile rapidity $Y$ with the collinearly improved kernel in Eq. (\ref{eq_kernel_p}) (left) and the BK equation in target rapidity $\eta$ with the collinearly improved kernel in Eq. (\ref{eq_kernel_t}) (right).
}
\label{fig_2dbk}
\end{figure}

An example of the 2D solution is given in the left part of Figure \ref{fig_2dbk}, wehre the BK equation was solved using the collinearly improved kernel
\begin{align}
    K(\Vec{r}, \Vec{r}_1, \Vec{r}_2) = \frac{\bar{\alpha}_s}{2\pi} \frac{r^2}{r_1^2r_2^2} \bigg[ \frac{r^2}{\min \lbrace r_1^2, r_2^2 \rbrace} \bigg]^{\pm \alpha_s A_1} \frac{J_1\left(2\sqrt{\bar{\alpha}_s \ln{\frac{r_1^2}{r^2}}\ln{\frac{r_2^2}{r^2}}}\right)}{\sqrt{\bar{\alpha}_s \ln{\frac{r_1^2}{r^2}}\ln{\frac{r_2^2}{r^2}}}}.
    \label{eq_kernel_p}
\end{align}
Here $\bar{\alpha}_s = \frac{N_C}{\pi} \alpha_s$ with \mbox{$\alpha_s = \alpha_s(\min\lbrace r, r_1, r_2 \rbrace)$} being the running strong coupling constant as described in \cite{Cepila:2019fiz} and $N_C = 3$ being the number of QCD colours. The constant $A_1 = \frac{11}{12}$ and $J_1$ is the Bessel function. 

The initial condition for this solution was proposed in \cite{Cepila:2019fiz}  and reads
\begin{align}
    N(Y = 0, r, b) = 1 - \exp{\left[-\frac{Q_s^2}{4}r^2 \exp{\left(-\frac{b^2}{2B} - \frac{r^2}{8B}\right)}\right]},
\end{align}
with the process scale $Q_s^2 = 0.496 \mathrm{~GeV}^2$ and parameter $B = 3.2258 \mathrm{~GeV}^{-2}$. 

\begin{figure}[!ht]
\centering
\includegraphics[width=.49\linewidth]{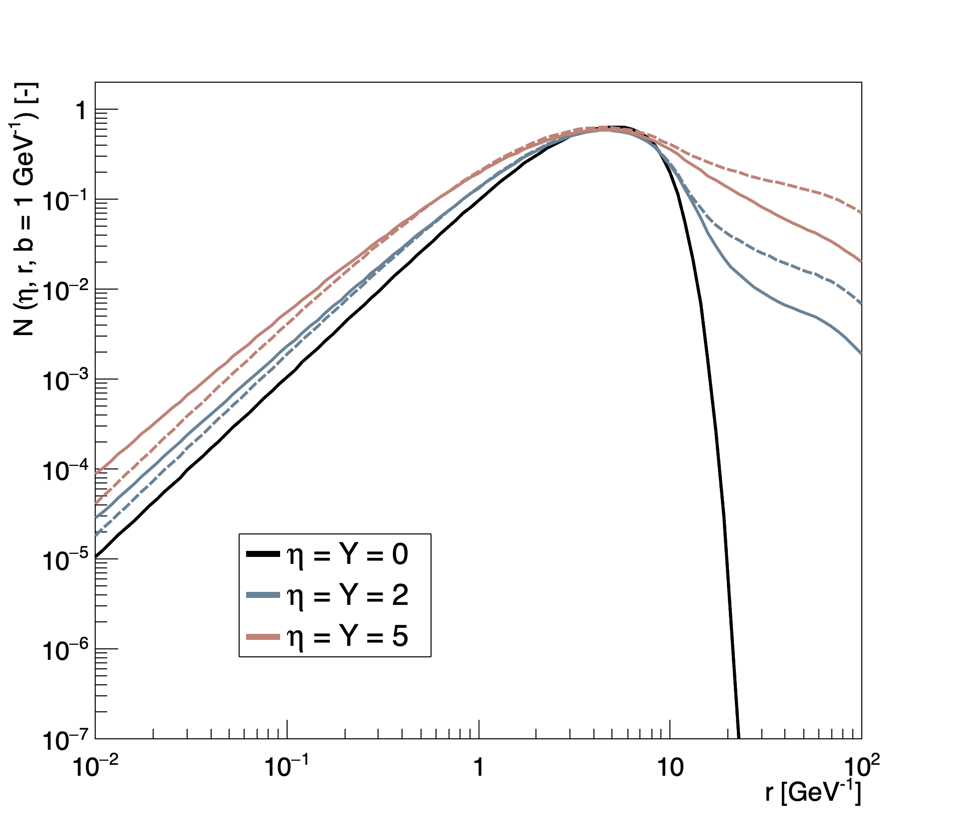}
\includegraphics[width=.49\linewidth]{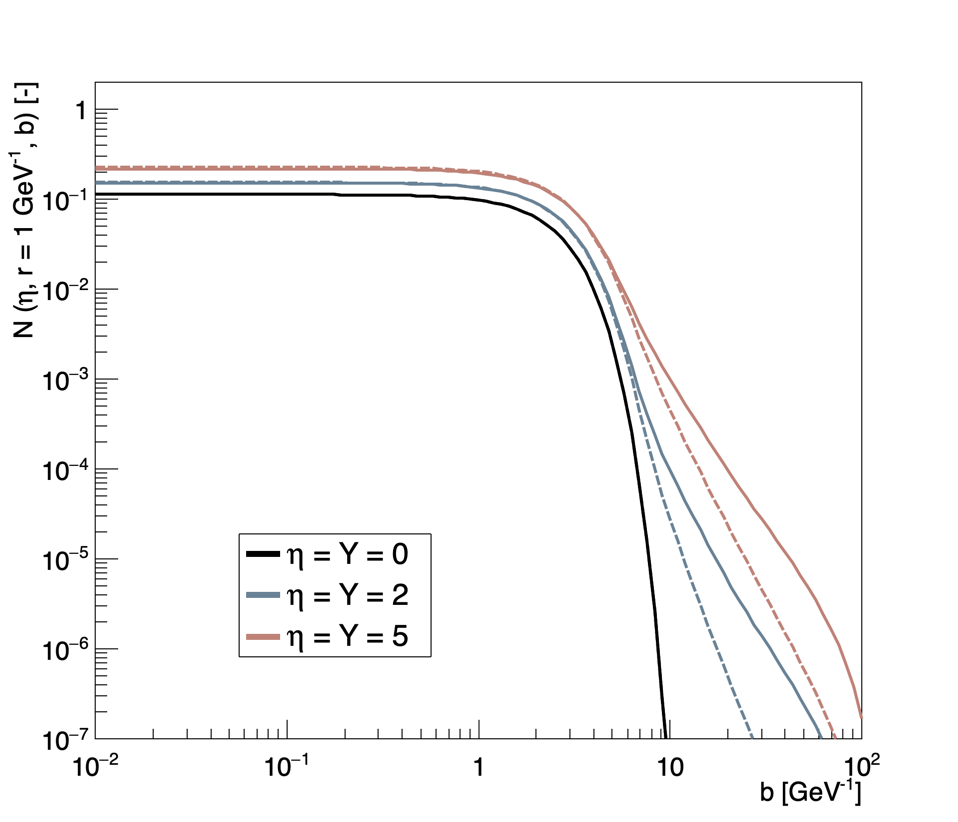}
\caption{
One-dimensional cuts of the dipole scattering amplitude at fixed impact parameter $b = 1 \mathrm{GeV}^{-1}$ (left) and fixed dipole size $r = 1 \mathrm{GeV}^{-1}$ (right) at initial condition (black curves) and subsequent steps of evolution. The evolution was driven by the collinearly improved kernel in Eq. (\ref{eq_kernel_t}) for the target rapidity (full curves) and in Eq. (\ref{eq_kernel_p}) for the projectile rapidity (dashed curves).
}
\label{fig_cuts}
\end{figure} 

%%%%%%%%%%%%%%%%%%%%%%%%%%%%%%%%%%%%%%%%%%

\section{BK equation in target rapidity}

Recently, it was pointed out (see Ref.~\cite{Ducloue:2019ezk}) that a slightly different formulation of the BK equation should better describe the nature of the interaction. Namely, there are two modifications that should be made. 

First, the BK equation should be reformulated in terms of the rapidity of the proton (target rapidity), defined as 
\begin{align}
    \eta = \ln{\frac{x_0}{x}},
\end{align}
to take the form 
\begin{align}
\partial_\eta N(\vec{r}, \vec{b}, \eta) = \int \mathrm{d} \vec{r}_1 K(r, r_1, r_2) \Big[ 	&   N(\vec{r}_1, \vec{b}_1, \eta_1) + N(\vec{r}_2, \vec{b}_2, \eta_2) - N(\vec{r}, \vec{b}, \eta)	\nonumber \\
		& - N(\vec{r}_1, \vec{b}_1, \eta_1) N(\vec{r}_2, \vec{b}_2, \eta_2) \Big], \label{eq_bk_t}
\end{align}
with the rapidities $\eta_j = \eta - \max\lbrace 0, \ln{\frac{r^2}{r^2_j}}\rbrace$. The collinearly improved kernel gets modified to 
\begin{align}
    K(\Vec{r}, \Vec{r}_1, \Vec{r}_2) = \frac{\bar{\alpha}_s}{2\pi} \frac{r^2}{r_1^2r_2^2} \bigg[ \frac{r^2}{\min \lbrace r_1^2, r_2^2 \rbrace} \bigg]^{\pm \alpha_s A_1}. \label{eq_kernel_t}
\end{align}

Second, the projectile rapidity needs to be redefined as 
\begin{align}
    Y = \eta + \ln{\frac{Q^2}{Q_s^2}},
\end{align}
where $Q^2$ is the virtuality and $Q_s^2$ is the scale of the process.

This formulation would then correspond better to the previous experience with solving the BFKL equation 
% \cite{fadin1995181, camici1997396, ciafaloni1998349} 
and should exhibit better behaviour in terms of both physics and numerics \cite{Ducloue:2019ezk}. To compare both formulations, a solution of the target rapidity BK equation, given by Eq. (\ref{eq_bk_t}), is shown in the right plot of Figure \ref{fig_2dbk}. For a better quantitative comparison, one-dimensional cuts of the dipole scattering amplitude are shown in Figure \ref{fig_cuts} at a fixed impact parameter $b$ (left) and a fixed dipole size $r$ (right).

%%%%%%%%%%%%%%%%%%%%%%%%%%%%%%%%%%%%%%%%%%

\section{Effect on observables}

\begin{figure}[!ht]
\centering
\includegraphics[width=\linewidth]{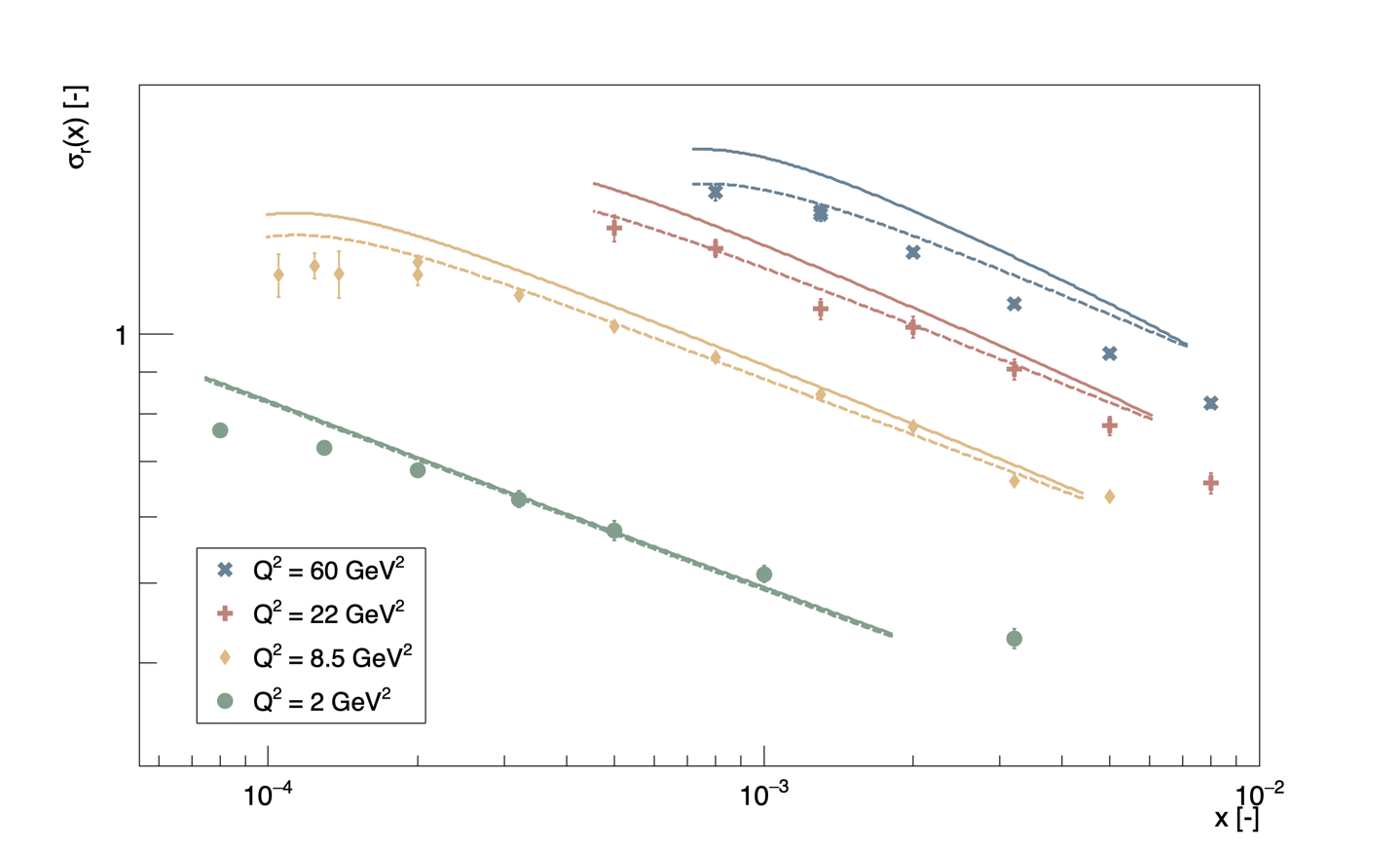}
\caption{Reduced cross section of the electron-proton scattering predicted by models using both target (full) and projectile (dashed) rapidity BK equations compared with data from HERA \cite{H1:2015ubc, Habib:2010zz}. Plots are shown at various fixed values of virtuality $Q^2$.}
\label{fig_sigma_r}
\end{figure}

To explore the effect the rapidity change has on the final results of the model, predictions of the electron-proton reduced cross  section,
\begin{align}
    \sigma_r (Q^2, x, y) = F_2(Q^2, x) - \frac{y^2}{1 + (1-y)^2} F_L(Q^2, x),
\label{eq_sigma_r}
\end{align}
are shown for both projectile and target rapidity formulations of the BK equation in comparison with experimental data in Figure \ref{fig_sigma_r}. The functions $F_2(Q^2, x)$ and $F_L(Q^2, x)$ in Eq. (\ref{eq_sigma_r}) are the structure functions and $y$ is the inelasticity.

As shown in Figure \ref{fig_sigma_r}, the transition from projectile to target rapidity in the formulation of the BK equation seems not to present any crucial numerical difference at the level of observable quantities. The fact that the target rapidity model exhibits overall a slightly worse data description comes naturally as the same numerical parameters ($Q_s^2, B$, etc, see \cite{Cepila:2019fiz}) are used in both models, although they were fitted for the projectile rapidity one. 

At higher virtualities (e.g. $Q^2 = 60 \mathrm{~GeV}^2$), the slope of the full (target rapidity) curves even seems to follow the data better, giving a good promise for future target rapidity BK studies.

%%%%%%%%%%%%%%%%%%%%%%%%%%%%%%%%%%%%%%%%%%

\section{Summary}
The parton-saturation-implementing BK evolution equation was solved in two different formulations: in terms of the target and the projectile rapidity. The resulting evolution of the dipole scattering amplitude is shown for both cases as two-dimensional graphs in Figure \ref{fig_2dbk} with one-dimensional cuts in Figure \ref{fig_cuts} for better quantitative comparison.

Both cases were used to calculate predictions of observables and an example of the reduced electron-proton cross section is shown in Figure \ref{fig_sigma_r} to provide a qualitative comparison. The transition from projectile to target rapidity does not seem to present any crucial numerical difficulties and the data description is satisfactory considering that for a clear comparison the same model parameters were used in both cases, although they were fitted for the projectile rapidity case.

%%%%%%%%%%%%%%%%%%%%%%%%%%%%%%%%%%%%%%%%%%

\vspace{6pt} 

%%%%%%%%%%%%%%%%%%%%%%%%%%%%%%%%%%%%%%%%%%

\section*{Acknowledgements}
We thank Dagmar Bendova for useful discussion. The work has been partially supported by the grant 22-27262S of the Czech Science Foundation (GACR).
The work was partially supported from European Regional Development Fund-Project "Center of Advanced Applied Science" No. CZ.02.1.01/0.0/0.0/16-019/0000778.

%%%%%%%%%%%%%%%%%%%%%%%%%%%%%%%%%%%%%%%%%%

\bibliography{literature}

\end{document}